\documentclass{emulateapj}
\usepackage{apjfonts}
\usepackage{color}

\newcommand{\pivec}{\mbox{\boldmath $\pi$}}
\newcommand{\muvec}{\mbox{\boldmath $\mu$}}

\definecolor{darkbrown}{RGB}{139,69,19}

\lefthead{HAN ET AL.} 
\righthead{OGLE-2015-BLG-0479A,B}

\begin{document}

\title{OGLE-2015-BLG-0479LA,B: Binary Gravitational Microlens Characterized by 
Simultaneous Ground-based and Space-based Observations}

\author{
C.~Han$^{1}$, A.~Udalski$^{2,31}$, A.~Gould$^{3,4,32,33}$, Wei~Zhu$^{3,33}$, R.~A.~Street$^{5,34,35}$ \\
and \\
J.~C.~Yee$^{6,36}$, C.~Beichman$^{7}$, C.~Bryden$^{11}$, S.~Calchi~Novati$^{7,8,9,37}$, S.~Carey$^{10}$, 
M.~Fausnaugh$^3$, B.~S.~Gaudi$^{3}$, Calen~B.~Henderson$^{11,38}$, Y.~Shvartzvald$^{11,38}$, B.~Wibking$^3$ \\
(The \textit{Spitzer} Microlensing Team),\\
M.~K.~Szyma{\'n}ski$^{2}$, I.~Soszy{\'n}ski$^{2}$, J.~Skowron$^{2}$, P.~Mr{\'o}z$^{2}$, 
R.~Poleski$^{2,3}$, P.~Pietrukowicz$^{2}$, S.~Koz{\l}owski$^{2}$, K.~Ulaczyk$^{2}$, 
{\L}.~Wyrzykowski$^{2}$, M.~Pawlak$^{2}$\\
(The OGLE Collaboration),\\
Y.~Tsapras$^{12}$, M.~Hundertmark$^{13}$, E.~Bachelet$^{14,15}$,
M.~Dominik$^{16,39}$, D.~M.~Bramich$^{15}$, A.~Cassan$^{17}$,
R.~Figuera~Jaimes$^{16,12}$, K.~Horne$^{16}$, C.~Ranc$^{17}$,
R.~Schmidt$^{12}$, C.~Snodgrass$^{18}$, J.~Wambsganss$^{12}$, 
I.~A.~Steele$^{19}$, J.~Menzies$^{20}$, S.~Mao$^{21}$\\
(The RoboNet collaboration),\\
V.~Bozza$^{8,9}$, U.~G.~J{\o}rgensen$^{22}$, K.~A.~Alsubai$^{15}$,
S.~Ciceri$^{4}$, G.~D'Ago$^{8,4,R}$, T.~Haugb{\o}lle$^{22}$, F.~V.~Hessman$^{23}$,
T.~C.~Hinse$^{24}$, D.~Juncher$^{22}$, H.~Korhonen$^{22,25}$, L.~Mancini$^{4}$,
A.~Popovas$^{22}$, M.~Rabus$^{4,26}$, S.~Rahvar$^{27}$, G.~Scarpetta$^{8,9,4}$,
J.~Skottfelt$^{22}$, J.~Southworth$^{28}$, D.~Starkey$^{16}$,
J.~Surdej$^{29}$, O.~Wertz$^{29}$, M.~Zarucki$^{9}$\\
(The MiNDSTEp consortium)\\
R.~W.~Pogge$^{3}$, D.~L.~DePoy$^{30}$\\
(The $\mu$FUN Collaboration).\\
}

\affil{$^{1}$  Department of Physics, Chungbuk National University, Cheongju 361-763, Republic of Korea}
\affil{$^{2}$ Warsaw University Observatory, Al. Ujazdowskie 4, 00-478 Warszawa, Poland}
\affil{$^{3}$ Department of Astronomy, Ohio State University, 140 W. 18th Ave., Columbus, OH 43210, USA}
\affil{$^{4}$ Max Planck Institute for Astronomy, K{\"o}nigstuhl 17, D-69117 Heidelberg, Germany}
\affil{$^{5}$ School of Physics and Astronomy, Queen Mary University of London, Mile End Road, London E1 4NS, UK}
\affil{$^{6}$ Harvard-Smithsonian Center for Astrophysics, 60 Garden St., Cambridge, MA 02138, USA}
\affil{$^{7}$ NASA Exoplanet Science Institute, MS 100-22, California Institute of Technology, Pasadena, CA 91125, USA}
\affil{$^{8}$ Dipartimento di Fisica "E.~R.~Caianiello", U\'niversit\'a di Salerno, Via Giovanni Paolo II, I-84084 Fisciano (SA), Italy}
\affil{$^{9}$ Istituto Internazionale per gli Alti Studi Scientifici (IIASS), Via G. Pellegrino 19, I-84019 Vietri Sul Mare (SA), Italy}
\affil{$^{10}$ Spitzer Science Center, MS 220-6, California Institute of Technology, Pasadena, CA, USA}
\affil{$^{11}$ Jet Propulsion Laboratory, California Institute of Technology, 4800 Oak Grove Drive, Pasadena, CA 91109, USA}
\affil{$^{12}$ Astronomisches Rechen-Institut, Zentrum f{\"u}r Astronomie der Universit{\"a}t Heidelberg (ZAH), 69120 Heidelberg, Germany}
\affil{$^{13}$ Niels Bohr Institute \& Centre for Star and Planet Formation, University of Copenhagen, {\O}ster Voldgade 5, 1350 - Copenhagen K, Denmark}
\affil{$^{14}$ Las Cumbres Observatory Global Telescope Network, 6740 Cortona Drive, suite 102, Goleta, CA 93117, USA}
\affil{$^{15}$ Qatar Environment and Energy Research Institute(QEERI), HBKU, Qatar Foundation, Doha, Qatar}
\affil{$^{16}$ SUPA, School of Physics \& Astronomy, University of St Andrews, North Haugh, St Andrews KY16 9SS, UK}
\affil{$^{17}$ Sorbonne Universit\'es, UPMC Univ Paris 6 et CNRS, UMR 7095, Institut d'Astrophysique de Paris, 98 bis bd Arago, 75014 Paris, France}
\affil{$^{18}$ Planetary and Space Sciences, Department of Physical Sciences, The Open University, Milton Keynes, MK7 6AA, UK}
\affil{$^{19}$ Astrophysics Research Institute, Liverpool John Moores University, Liverpool CH41 1LD, UK}
\affil{$^{20}$ South African Astronomical Observatory, PO Box 9, Observatory 7935, South Africa}
\affil{$^{21}$ National Astronomical Observatories, Chinese Academy of Sciences, 100012 Beijing, China}
\affil{$^{22}$ Niels Bohr Institutet, K{\o}benhavns Universitet, Juliane Maries Vej 30, DK-2100 K{\o}benhavn {\O}, Denmark}
\affil{$^{23}$ Institut f{\"u}r Astrophysik, Georg-August-Universit{\"a}t G{\"o}ttingen, Friedrich-Hund-Platz 1, D-37077 G{\"o}ttingen, Germany}
\affil{$^{24}$ Korea Astronomy and Space Science Institute, 776 Daedeokdae-ro, Yuseong-gu, 305-348 Daejeon, Republic of Korea}
\affil{$^{25}$ Finnish Centre for Astronomy with ESO (FINCA), University of Turku, V{\"a}is{\"a}l{\"a}ntie 20, FI-21500 Piikki{\"o}, Finland}
\affil{$^{26}$ Instituto de Astrof{\'i}sica, Facultad de F??sica, Pontificia Universidad Cat{`o}lica de Chile, Av.\ Vicu\~{n}a Mackenna 4860, 7820436 Macul, Santiago, Chile}
\affil{$^{27}$ Department of Physics, Sharif University of Technology, P.O. Box 11155-9161 Tehran, Iran}
\affil{$^{28}$ Astrophysics Group, Keele University, Staffordshire, ST5 5BG, UK}
\affil{$^{29}$ Institut d'Astrophysique et de G{\'e}ophysique, Universit{\'e} de Li{\`e}ge, 4000 Li{\`e}ge, Belgium}
\affil{$^{30}$ Department of Physics and Astronomy, Texas A\&M University, College Station, TX 77843-4242, USA}

\footnotetext[31]{The OGLE Collaboration.}
\footnotetext[32]{The $\mu$FUN collaboration.}
\footnotetext[33]{The $Spitzer$ Microlensing Team.}
\footnotetext[34]{The RoboNet collaboration.}
\footnotetext[35]{The MiNDSTEp consortium.}
\footnotetext[36]{Sagan Fellow.}
\footnotetext[37]{Sagan Visiting Fellow.}
\footnotetext[38]{NASA Postdoctoral Program Fellow.}
\footnotetext[39]{Royal Society University Research Fellow.}

\begin{abstract}

We present a combined analysis of the observations of the
gravitational microlensing event OGLE-2015-BLG-0479 taken both from
the ground and by the {\it Spitzer Space Telescope}.  The light curves
seen from the ground and from space exhibit a time offset of $\sim 13$
days between the caustic spikes, indicating that the relative
lens-source positions seen from the two places are displaced by
parallax effects.  From modeling the light curves, we measure the
space-based microlens parallax.  Combined with the angular Einstein
radius measured by analyzing the caustic crossings, we determine the
mass and distance of the lens.  We find that the lens is a binary
composed of two G-type stars with masses $\sim 1.0\ M_\odot$ and $\sim
0.9\ M_\odot$ located at a distance $\sim 3$ kpc.  In addition, we are
able to constrain the complete orbital parameters of the lens thanks
to the precise measurement of the microlens parallax derived from the
joint analysis.  In contrast to the binary event OGLE-2014-BLG-1050,
which was also observed by {\it Spitzer}, we find that the
interpretation of OGLE-2015-BLG-0479 does not suffer from the degeneracy
between $(\pm,\pm)$ and $(\pm,\mp)$ solutions, confirming that the
four-fold parallax degeneracy in single-lens events collapses into the
two-fold degeneracy for the general case of binary-lens events.  The
location of the blend in the color-magnitude diagram is consistent
with the lens properties, suggesting that the blend is the lens itself.
The blend is bright enough for spectroscopy and thus this possibility
can be checked from future follow-up observations.

\end{abstract}


\section{INTRODUCTION}

Einstein radii of typical Galactic gravitational microlensing events
are of order AU.  Hence, if lensing events are observed from a
satellite in a solar orbit, the relative lens-source positions seen
from the ground and from the satellite appear to be different,
resulting in different light curves. Combined analysis of the light
curves observed both from the ground and from the satellite leads to
the measurement of the microlens parallax vector $\pivec_{\rm E}$
\citep{Refsdal1966, Gould1994}, which is referred to as the
``space-based microlens parallax''. Measurement of $\pivec_{\rm E}$ is
important because it enables one to constrain the mass $M$ and
distance $D_{\rm L}$ to the lensing object by
\begin{equation}
M={\theta_{\rm E}\over \kappa \pi_{\rm E}};\qquad
D_{\rm L}={{\rm AU}\over \pi_{\rm E}\theta_{\rm E}+\pi_{\rm S}},
\label{eq1}
\end{equation}
where $\theta_{\rm E}$ is the angular Einstein radius, $\kappa=4G/(c^2
{\rm AU})$, $\pi_{\rm S}={\rm AU}/D_{\rm S}$ is the parallax of the
lensed star (source), and $D_{\rm S}$ is the distance to the source.
Microlens parallaxes can be measured from the single platform of Earth
that is being accelerated by its orbital motion around the Sun.
Although parallaxes of most lenses with known physical parameters were
measured in this way, ground-based measurement of microlens
parallaxes, referred to as annual microlens parallaxes, has limited
applicability, primarily to the small fraction of long time-scale
events caused by nearby lenses.  Therefore, space-based microlens
parallax provides the only way to routinely measure microlens
parallaxes for an important fraction of microlensing events.

In 2014, the 50-year old concept of the space-based microlens parallax
measurement was realized by a microlensing program making use of the
{\it Spitzer Space Telescope} \citep{Gould2014}, which has a projected
separation from the Earth $\sim 1$ AU. The principal goal of the
program is determining the Galactic distribution of planets by
measuring microlens parallaxes and thereby estimating distances of the
individual lenses \citep{Calchi2015a}.  From combined observations
both from the ground and from the {\it Spitzer} telescope conducted in
2014 and 2015 seasons, the masses and distances of two microlensing
planets were successfully determined \citep{Udalski2015b,Street2016}.

Besides planetary microlensing events, another important target
lensing events of {\it Spitzer} observations are those produced by
binary objects, especially caustic-crossing binary-lens events.
Caustics in gravitational lensing phenomena refer to the positions on
the source plane at which a point-source would be infinitely
magnified. In reality, source stars have finite sizes and thus
lensing magnifications during caustic crossings deviate from those of
a point source. Detecting these finite-source effects enables one to
measure the angular Einstein radius $\theta_{\rm E}$, which is the
other ingredient needed for the unique determinations of $M$ and
$D_{\rm L}$ (see Equation~\ref{eq1}).  The usefulness of {\it Spitzer}
observations in characterizing binaries was demonstrated by the
microlens parallax measurements for two caustic-crossing binary-lens
events \citep{Zhu2015, Shvartzvald2015}.

In this paper, we present the analysis of the caustic-crossing
binary-lens event OGLE-2015-BLG-0479, which was simultaneously
observed by ground-based telescopes and the {\it Spitzer Space
  Telescope} in the 2015 season. By measuring both the lens parallax
and the angular Einstein radius, we are able to determine the mass and
distance to the lens.  In addition, we can constrain the complete
orbital parameters of the lens thanks to the precisely measured
microlens parallax by the {\it Spitzer} data.  We also investigate
modeling degeneracies by comparing the event with OGLE-2014-BLG-1050
\citep{Zhu2015}, which is another caustic-crossing binary-lens event
observed by {\it Spitzer} with similar photometric precision, cadence,
and coverage.

\begin{figure*}[t]
\epsscale{0.7}
\plotone{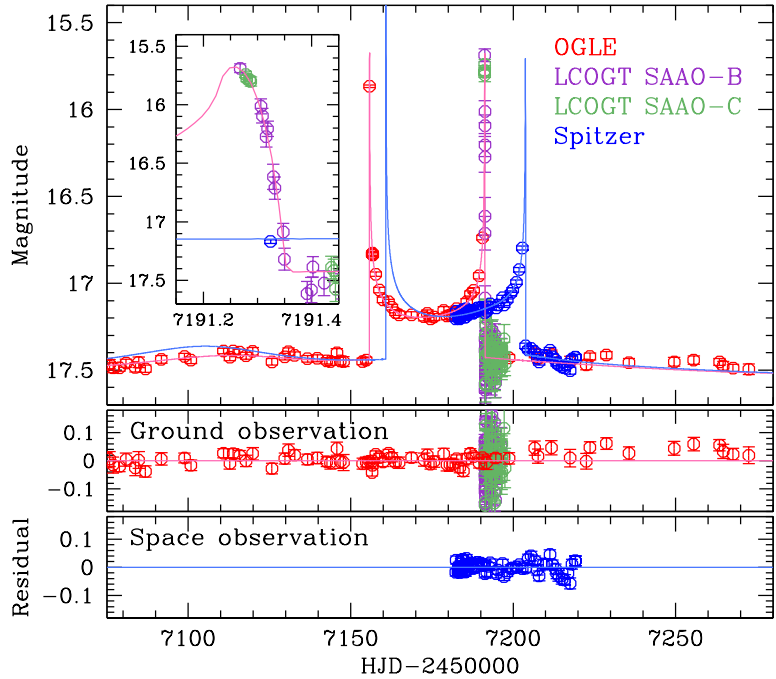}
\caption{\label{fig:one}
Light curves of the microlensing event OGLE-2015-BLG-0479 as seen from
the Earth and from the {\it Spitzer} telescope.  Superposed on the
data points are the best-fit model curves obtained considering
space-based parallax effects.  The insets shows an enlargement of the
caustic-exit part of the light curve seen from Earth.  The two lower
panels show the residuals from the model for the ground-based and
space-based data sets.
}
\end{figure*}

\section{OBSERVATION}

The event OGLE-2015-BLG-0479 occurred on a star located in the
Galactic bulge field with coordinates $({\rm RA},{\rm DEC})_{\rm
  J2000}= (17^\circ 43'40''\hskip-2pt.6, -35^{\rm h}30^{\rm m}33^{\rm
  s}\hskip-2pt.4)$, that corresponds to the Galactic coordinates
$(l,b)=(354.18^\circ,-3.08^\circ)$.  It was discovered by the Early
Warning System \citep[EWS:][]{Udalski2015a} of the OGLE group on 2015
March 18 (${\rm HJD}'={\rm HJD}-2450000 \sim 7100)$ from survey
observations conducted using the 1.3m telescope located at Las
Campanas Observatory in Chile.

On 2015 May 13 (${\rm HJD}'\sim 7155.5)$, the event exhibited a sharp
rise of the source brightness and the onset of this anomaly was
announced to the microlensing community.  Such a rise in the light
curve is a characteristic feature that occurs when a source star
enters a caustic formed by a binary object.  In response to the
anomaly alert, the $\mu$FUN collaboration \citep{Gould2006} conducted
follow-up observations using the 1.0m telescope at Cerro Tololo
Inter-American Observatory (CTIO) in Chile.  After the sharp rise, the
light curve exhibited a ``U''-shape brightness variation, which is a
characteristic feature when the source moves inside of a binary
caustic.  Caustics produced by binary lenses are closed curves, and
thus a caustic exit was anticipated.  On ${\rm HJD}'\sim 7191$, the
source brightness suddenly dropped, indicating that the source exited
the caustic.  The RoboNet collaboration and the MiNDSTEp consortium,
who were watching the progress of the event, conducted intensive
observations during the caustic exit using two 1.0m telescopes of Las
Cumbres Observatory Global Telescope Network (LCOGT) located in South
African Astronomical Observatory (SAAO).  Thanks to the follow-up
observations, the caustic exit was densely resolved.

The event was also observed from space as a part of the {\it Spitzer}
microlensing program.  The general description of the program and
target selection protocol in 2015 season are given in
\citet{Udalski2015b} and \citet{Yee2015}, respectively.  {\it Spitzer}
observations were conducted for 37 days from 2015 June 8 (${\rm
  HJD}'\sim 7182$) to July 15 (${\rm HJD}'\sim 7219$).  The event was
observed with a half-day cadence until June 18 (${\rm HJD}'\sim
7192$), just after the caustic exit seen from the ground, and one-day
cadence thereafter. From these observations, a total of 59 data points
were obtained.

Data from ground-based observations were processed using pipelines
that are based on the Difference Image Analysis method
\citep{Alard1998, Wozniak2000} and customized by the individual groups
\citep{Udalski2003, Bramich2008}.  Data from {\it Spitzer}
observations were processed by using a photometry algorithm that is
optimized for images taken by the Infrared Array Camera of {\it
  Spitzer} in crowded fields \citep{Calchi2015b}.

In Figure~\ref{fig:one}, we present the light curve of
OGLE-2015-BLG-0479.  
One finds that both light curves observed from the ground and from the
{\it Spitzer} telescope are characterized by distinctive
caustic-crossing features.  We note that both the caustic entrance and
exit were captured by the ground-based data, while only the caustic
exit was captured by the space-based data.  The light curves observed
from the ground and from the {\it Spitzer} telescope exhibit a $\sim
13$ day offset between the times of the caustic exits, indicating that
the relative lens-source positions are displaced by the parallax
effect.

Another important characteristics of the light curves is that the
duration between the caustic crossings in the ground-based light
curve, $\sim ~35$ days, comprises a significant fraction of the whole
duration of the event ($\sim 180$ days). This indicates that the
source is likely to have crossed a big caustic formed by a binary lens
with roughly equal mass components and a separation similar to the
Einstein radius corresponding to the total mass of the lens.  This is
further evidenced by the fact that the space-based light curve also
exhibits a strong caustic-crossing feature that could not have been
produced if the caustic were small compared to the displacement of the
source trajectory by parallax effects.

In many respects, OGLE-2015-BLG-0479 is similar to OGLE-2014-BLG-1050
\citep{Zhu2015}, which is another caustic-crossing binary-lens event
simultaneously observed from the ground and from the {\it Spitzer}
telescope.  First, the light curves of both events exhibit distinctive
caustic-crossing features with wide time gaps between the
caustic-crossing spikes.  Second, the {\it Spitzer} data cover the
caustic exit but miss the entrance for both events.  Third, both
events have similar time scales and were covered with similar
photometric precision and cadence.  Hence, it will be interesting to
compare the results of analysis, particularly regarding the four-fold
degeneracy that was identified to exist for OGLE-2014-BLG-1050.  See
Section 3 for more details about the degeneracy.

\section{MODELING}

Light curves of single-mass lensing events obtained from both space-
and ground-based observations yield four sets of degenerate solutions
\citep{Refsdal1966} , which are often denoted by $(+,+)$, $(-,-)$,
$(+,-)$, and $(-,+)$, where the former and latter signs in each
parenthesis represent the signs of the lens-source impact parameters
as seen from Earth and from the satellite, respectively. This
four-fold degeneracy occurs due to the fact that a pair of light
curves resulting from the source trajectories seen from Earth and from
the satellite passing on the same side with respect to the lens,
i.e. $(+,+)$ or $(-,-)$ solutions, are similar to the pair of light
curves resulting from source trajectories passing on the opposite
sides of the lens, i.e. $(+,-)$ or $(-,+)$ solutions. For the
graphical presentation of the four-fold degeneracy, see Figure 2 of
\citet{Gould1994}.

For well covered binary-lens events, it is expected that the
degeneracy between the pair of $(+,+)$ and $(+,-)$ [or $(-,-)$ and
  $(-,+)$] solutions are generally resolved due to the lack of lensing
magnification symmetry compared to the single lens case. The remaining
degeneracy, i.e. $(+,+)$ versus $(-,-)$, may persists, but these
solutions usually give similar amplitudes of the microlens parallax,
and thus the physical lens parameters estimated from the two
degenerate solutions are similar to one another. In the case of
OGLE-2014-BLG-1050, \citet{Zhu2015} found that the four-fold
degeneracy unexpectedly persisted and diagnosed that the degeneracy
remained unresolved because (1) {\it Spitzer} data partially covered
the light curve and (2) the source-lens relative motion happened to be
almost parallel to the direction of the binary-lens axis. Similar to
OGLE-2014-BLG-1050, the {\it Spitzer} data of OGLE-2015-BLG-0479 cover
only the caustic exit of the light curve, and thus the degeneracy may
persist. We, therefore, investigate the possibility of the degeneracy.

Modeling the light curve of OGLE-2015-BLG-0479 is carried out in multiple steps:
\begin{enumerate}
\item
preliminary modeling based on the ground-based data,
\item
measuring the microlens parallax with combined ground- and space-based
data, and
\item
refining the identified solutions.
\end{enumerate}
In the following paragraphs, we describe these in detail.

In the first step, we conduct a preliminary modeling of the light
curve obtained from ground-based observations in order to find an
initial position in the parameter space from which $\chi^2$
minimization can be initiated.  This preliminary modeling is based on
the 7 principal binary lensing parameters plus 2 flux parameters for
the data set obtained by each telescope.  The first four of these
principal parameters describe the lens-source approach, including
$t_0$, $u_0$, $t_{\rm E}$, and $\alpha$, where $t_0$ is the time of
the closest source approach to a reference position of the lens, $u_0$
is the source-reference separation at $t_0$ (impact parameter),
$t_{\rm E}$ is the time scale for the source to cross the angular
Einstein radius $\theta_{\rm E}$ of the lens (Einstein time scale),
and $\alpha$ is the angle between the source trajectory and the binary
axis (source trajectory angle).  We choose the center of mass of the
binary lens as the reference position.  Another 2 principal parameters
characterize the binary lens including $s_\perp$ and $q$, where
$s_\perp$ is the projected separation and $q$ is the mass ratio
between the binary lens components. We note that the parameters $u_0$
and $s_\perp$ are normalized to $\theta_{\rm E}$. The last parameter
$\rho$, which is defined as the ratio of the angular source radius
$\theta_*$ to the Einstein radius, i.e., $\rho=\theta_*/\theta_{\rm
  E}$ (normalized source radius), is needed to account for the
caustic-crossing parts of the light curve affected by finite-source
effects.  The two flux parameters $F_s$ and $F_b$ represent the fluxes
from the source and blended light, respectively.  The principal
lensing parameters are searched for by using a downhill approach based
on the Markov Chain Monte Carlo (MCMC) method. The flux parameters
$F_s$ and $F_b$ are searched for by a linear fitting.

Magnifications affected by finite-source effects are computed by using
a combination of numerical and semi-analytic methods. In the immediate
neighboring region around caustics, we use the numerical
inverse-ray-shooting method \citep{Schneider1986}. In the outer region
surrounding caustics, we use the semi-analytic hexadecapole
approximation \citep{Pejcha2009, Gould2008}.

In computing finite-source magnifications, we consider surface
brightness variation of the source star caused by limb darkening by
modeling the surface brightness profile as
\begin{equation}
S_\lambda \propto \left[ 1-\Gamma_\lambda \left(1-{3\over 2}\cos\phi \right)\right],
\label{eq2}
\end{equation}
where $\Gamma_\lambda$ is the linear limb-darkening coefficient and
$\phi$ is the angle between the normal to the source surface and the
line of sight toward the center of the source star.  The values of the
limb-darkening coefficient are chosen from the catalog of
\citet{Claret2000} based on the source type determined from the
de-reddened color and brightness.  We find that the source is an early
K-type subgiant and adopt $\Gamma_I=0.53$ and $\Gamma_L=0.22$.  For
the detailed procedure of determining the source type, see Section 4.

In the second step, we conduct another modeling including the {\it
  Spitzer} data and considering parallax effects, starting from the
solution found from the preliminary modeling. Parallax effects are
incorporated by two parameters $\pi_{{\rm E},N}$ and $\pi_{{\rm
    E},E}$, which are the two components of the lens parallax vector
$\pivec_{\rm E}$ projected onto the sky along the north and east
equatorial coordinates, respectively. The starting values of the lens
parallax parameters $\pi_{{\rm E},N}$ and $\pi_{{\rm E},E}$ can be, in
principle, estimated from the offsets in the values of $t_0$ and $u_0$
for the two light curves observed from the ground and from the {\it
  Spitzer} telescope because the parallax vector is related to these
offsets by
\begin{equation}
\pivec_{\rm E}={{\rm AU}\over D_\perp} \left( {\Delta t_0\over t_{\rm E}}, \Delta u_0\right),
\label{eq3}
\end{equation}
where $\Delta t_0=t_{0,{\rm sat}}-t_{0,\oplus}$, $\Delta u_0=u_{0,{\rm
    sat}}-u_{0,\oplus}$, and $D_\perp$ is the projected separation
between Earth and the satellite. During the time of the event,
$D_\perp \sim 1.4$ AU. However, this analytic estimation of the lens
parallax vector is difficult because $t_{0,{\rm sat}}$, and $u_{0,{\rm
    sat}}$ are uncertain due to the partial coverage of the event by
the {\it Spitzer} data.  Another way to obtain a starting $\pivec_{\rm
  E}$ value is conducting an additional modeling based on the
ground-based data but this time considering the annual parallax
effects, which affect the ground-based light curve via Earth's annual
orbital motion.  We find that implementing this method is also
difficult because the photometric data are not good enough and cadence
of ground-based observation is not high enough to precisely measure
$\pivec_{\rm E}$ based on subtle deviations caused by the annual
parallax. We therefore conduct a grid search in the $\pi_{{\rm E},N} -
\pi_{{\rm E},E}$ plane. In addition to finding a starting value of
$\pivec_{\rm E}$, this second-step grid search is needed to identify
possibly multiple solutions resulting from the parallax degeneracy.

In the final step, we identify local solutions found from the
second-step grid search and refine them by letting all parameters
vary. In this step, we additionally consider the effect of lens
orbital motion, which is known to induce long-term deviations in
binary-lensing light curves similar to the deviation induced by
parallax effects \citep{Park2013}.  Orbital effects cause the
projected binary separation $s_\perp$ and the source trajectory angle
$\alpha$ to vary in time.  Under the assumption that the orbital
period $P$ is much greater than the event time scale, i.e.\ $P \gg
t_{\rm E}$\footnote{A large binary-lens caustic forms when the
separation between the binary components is similar to the physical
Einstein radius $r_{\rm E}$, i.e., $a\sim r_{\rm E}$.  The Einstein
radius is related to the mass and distance to the lens by
$$
r_{\rm E}\sim 4~{\rm AU} \left( {M\over M_\odot}\right)
\left[ { x(1-x)\over 0.25}\right]^{1/2},
$$ 
where $x=D_{\rm L}/D_{\rm S}$ \citep{Gaudi2012}.  With the Kepler
law, $(P/{\rm yr})^2=(a/{\rm AU})^3/(M/M_\odot)$, the orbital period
is expressed as
$$
P \simeq 8\ {\rm yr} 
\left( {r_{\rm E}\over 4\ {\rm AU}}\right)^{3/2}
\left( {M\over M_\odot}\right)^{1/4}
\left[ { x(1-x)\over 0.25}\right]^{3/4}.
$$
Considering that a typical Einstein time scale 
$$
t_{\rm E}\simeq 35\ {\rm day} \left( {M\over M_\odot} \right)^{1/2},
$$
the orbital period of a binary lens is much greater than the Einstein
time scale, and thus the assumption $P \gg t_{\rm E}$ is valid in most
cases of Galactic binary-lens events.}, the variations of $s_\perp$
and $\alpha$ can be approximated to be linear and the lens-orbital
effect is described by two parameters $ds_\perp/dt$ and $d\alpha/dt$
that are the linear change rates of the projected binary separation
and the source trajectory angle, respectively.  For the full
consideration of the Kepler orbital motion, on the other hand, one
needs two additional parameters $s_\parallel$ and $ds_\parallel/dt$,
which represent the line-of-sight separation between the binary lens
components and its rate of change, respectively.  See
\citet{Skowron2011} for the full description of the orbital lensing
parameters.  In our analysis, we test both orbital models based on the
linear approximation with 2 parameters and the full Keplerian orbital
motion with 4 parameters.

\begin{figure}[t]
\epsscale{1.20}
\plotone{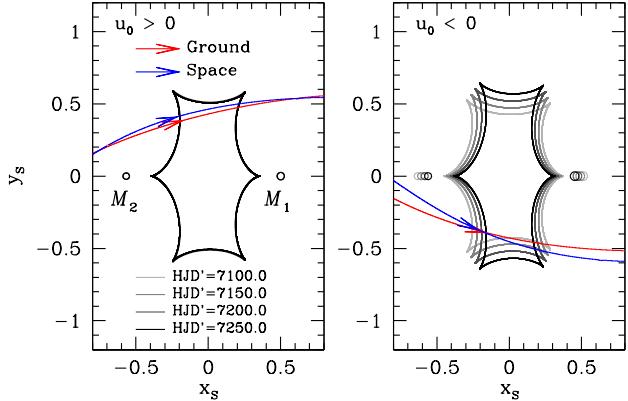}
\caption{\label{fig:two}
Geometry of the lens system for the $u_0>0$ (left panel) and $u_0<0$
(right panel) solutions.  In each panel, the closed curve with 6 cusps
represents the caustic formed by the binary lens.  The locations of
the binary lens components ($M_1$ and $M_2<M_1$) are marked by small
open circles.  The red and blue curves with arrows are the source
trajectories as seen from Earth and from the {\it Spitzer} telescope,
respectively.  We note that the positions of the lens components and
the shape of the caustic vary in time due to the orbital motion of the
binary lens.  We present the positions of the lens and caustic at 4
different times marked in the legend.  We note that the variation of
the caustic for the $u_0>0$ model is very small due to the small value
of the orbital parameter $ds_\perp/dt$ and thus 4 different caustics
appear to be a single caustic.  All lengths are normalized to the
angular Einstein radius corresponding to the total mass of the binary
lens.
}
\end{figure}

\section{Solutions}

In Table~\ref{table:one}, we present the lensing parameters of the
solutions found from modeling.  We present two sets of solutions with
$u_0>0$, i.e., $(+,+)$ solution, and $u_0<0$, i.e., $(-,-)$ solution,
because the degeneracy between the two solutions is very severe with
$\Delta\chi^2\sim 3.5$.  We note that the two degenerate solutions are
in mirror symmetry with respect to the binary axis and thus the
parameters of the solutions are in the relation $(u_0, \alpha,
\pi_{{\rm E},N}, d\alpha/dt) \leftrightarrow -(u_0, \alpha, \pi_{{\rm
    E},N}, d\alpha/dt)$.  The uncertainty of each parameter is
determined as the standard deviation of the distribution derived from
the MCMC chain.  In Figure~\ref{fig:one}, we present the best-fit
model light curve ($u_0<0$ solution) superposed on the observed
data. The model curves for the ground- and space-based data sets are
presented in different colors that are in accordance with those of the
individual data sets.  We find that the model based on the full
Keplerian orbital motion provides a better fit than the model based on
the linear approximation with $\Delta\chi^2\sim 30$.

Figure~\ref{fig:two} shows the geometry of the lens system, where the
left and right panels are for the $u_0>0$ and $u_0<0$ solutions,
respectively.  We note that the degeneracy between the $u_0>0$ and
$u_0<0$ solutions, which is referred to as the ``ecliptic degeneracy''
\citep{Skowron2011}, is known to exist for general binary-lens events.
In each panel, the red and blue curves with arrows represent the
source trajectories seen from the ground and from space, respectively,
and the closed curve with six cusps represents the caustic. We note
that the shape of the caustic varies in time due to the orbital motion
of the binary lens. From the geometry, it is found that the sharp
spikes were produced by the crossings of the source over the single
big caustic formed by a binary having a roughly equal mass ($q\sim
0.85$) components with a projected separation similar to the Einstein
radius ($s_\perp\sim 1.1$).  The source seen from the ground and from
space took different trajectories where the space-based source
trajectory trailed the ground-based trajectory with a time gap $\sim
13$ days and with a slightly different source trajectory angle.  The
weak bump at ${\rm HJD}'\sim 7115$ in the ground-based light curve was
produced when the source approached the cusp of the caustic located on
the binary axis close to the lower-mass binary component.  We find
that improvement of the fit with the consideration of the lens orbital
motion is $\Delta\chi^2\sim 43.5$.
In Appendix, we discuss the false alarm probabilities 
associated with the introduction of the additional orbital-motion parameters 
relative to the standard model.

\begin{figure*}[t]
\epsscale{0.85}
\plotone{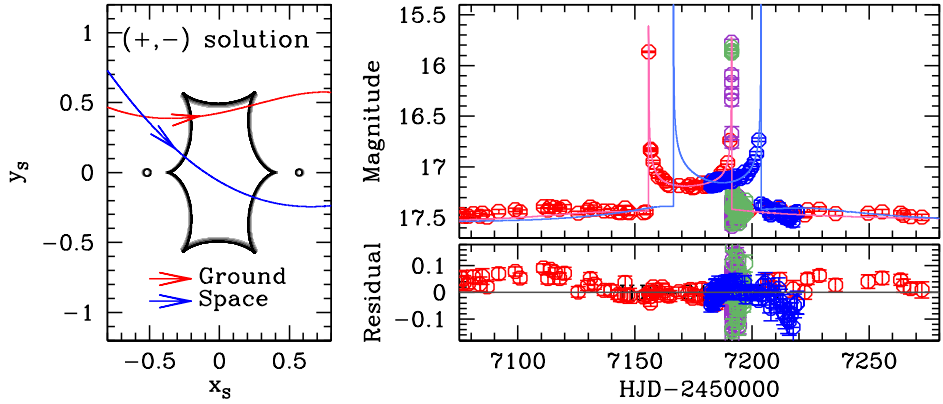}
\caption{\label{fig:three}
The lens system geometry and the light curve corresponding to the
$(+,-)$ local solution.  Notations are same as those in
Fig.~\ref{fig:one} and \ref{fig:two}.
}
\end{figure*}

Although the degeneracy between $u_0>0$ and $u_0<0$ solutions
persists, we find that the degeneracy between $(\pm,\pm)$ and
$(\pm,\mp)$ solutions is clearly resolved.  In Figure~\ref{fig:three},
we present the lens system geometry and the model light curve
corresponding to the $(+,-)$ local solution.  We find that although
the local solution explains the caustic-crossing features, the fit in
the wings of the light curve is poor with $\chi^2$ difference from the
$(-,-)$ solution $\Delta\chi^2=155$.  Therefore, the result confirms
that the four-fold degeneracy in single-lens events collapses into the
two-fold degeneracy in general binary-lens events.

\begin{deluxetable}{lrr}
\tablecaption{Lensing parameters\label{table:one}}
\tablewidth{0pt}
\tablehead{
\multicolumn{1}{c}{Parameters}     &
\multicolumn{1}{c}{$u_0>0$}        &
\multicolumn{1}{c}{$u_0<0$} 
}
\startdata
$\chi^2$                         &   736.4                 &  732.9                 \\     
$t_0 \ ({\rm HJD}-2450000)$      &  7163.992  $\pm$ 0.743   &   7166.439 $\pm$ 0.179   \\     
$u_0$                            &   0.417    $\pm$ 0.005   &  -0.418    $\pm$ 0.004   \\     
$t_{\rm E} \ ({\rm days})$       &  91.0      $\pm$ 1.7     &   86.3     $\pm$ 0.5     \\     
$s_\perp$                        &   1.07     $\pm$ 0.01    &   1.10     $\pm$ 0.01    \\     
$q$                              &   0.88     $\pm$ 0.05    &   0.81     $\pm$ 0.03    \\     
$\alpha \ ({\rm rad})$           &  -0.270    $\pm$ 0.017   &   0.242    $\pm$ 0.003   \\     
$\rho \ (10^{-3})$               &   0.75     $\pm$ 0.14    &   0.73     $\pm$ 0.12    \\ 
$\pi_{{\rm E},N}$                &   0.01     $\pm$ 0.01    &  -0.06     $\pm$ 0.01    \\              
$\pi_{{\rm E},E}$                &  -0.12     $\pm$ 0.01    &  -0.11     $\pm$ 0.01    \\              
$ds_\perp/dt$ $({\rm yr}^{-1})$  &   0.01     $\pm$ 0.08    &  -0.32     $\pm$ 0.05    \\
$d\alpha/dt$ $({\rm yr}^{-1})$   &   0.53     $\pm$ 0.04    &  -0.40     $\pm$ 0.01    \\
$s_\parallel$                    &   0.21     $\pm$ 0.28    &  -1.09     $\pm$ 0.19    \\
$ds_\parallel/dt$ $(yr^{-1})$    &   0.50     $\pm$ 0.35    &  -0.01     $\pm$ 0.25    
\enddata
\end{deluxetable}

\section{LENS PARAMETERS}

\subsection{Angular Einstein Radius}

In addition to the microlens parallax, one additionally needs to
estimate the angular Einstein radius in order to uniquely determine
the lens mass and distance.  The angular Einstein radius is measured
by analyzing the caustic-crossing parts of the light curve that are
affected by finite-source effects. This analysis yields the normalized
source radius $\rho=\theta_*/\theta_{\rm E}$. By deducing the angular
source radius $\theta_*$ from the de-reddened color and brightness,
the angular Einstein radius is determined by $\theta_{\rm
  E}=\theta_*/\rho$.

\begin{figure}[t]
\epsscale{1.15}
\plotone{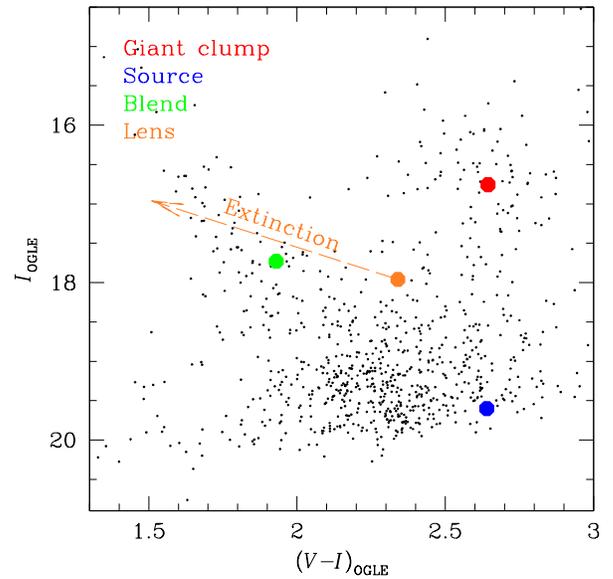}
\caption{\label{fig:four}
Position of the source star with respect to the centroid of giant
clump in the instrumental color-magnitude diagram.  Also presented are
the positions of the blend and the lens.  The lens position is
estimated under the assumption that the lens and clump giants
experience the same amount of reddening and extinction.  The arrow
starting from the lens position represents one magnitude difference in
extinction (relative to the clump), under the assumption that the
ratio of total-to-selective extinction is $R_{VI} = A_I/E(V-I)= 1.31$.
Hence, if the blend is the lens, then the lens is less extincted than
the clump by $\Delta A_I \simeq 0.5$.  \color{black}
}
\end{figure}

The de-reddened color $(V-I)_0$ and brightness $I_0$ of the source
star are estimated through multiple steps.  We first determine the
instrumental $I-H$ color based on the $\mu$FUN CTIO $I$ and $H$-band
data by linear regression of fluxes measured at various magnifications
during the event.  We then convert $I-H$ into $V-I$ using the
color-color relation of \citet{Bessell1988} and find that $V-I=0.87\pm
0.04$.  The instrumental $I$-band magnitude of the source star,
$I=19.6$, is estimated based on the $F_s$ and $F_b$ values determined
from modeling of the OGLE data.  Once the instrumental color $V-I$ and
brightness $I$ are determined, we then calibrate them based on the
relative position of the source star in the instrumental
color-magnitude diagram with respect to the centroid of giant clump
(GC), for which its de-reddened color and brightness are known to be
constant, $(V-I)_{0,{\rm GC}}=1.06$ \citep{Bensby2011} and $I_{0,{\rm
    GC}}=14.7$ \citep{Nataf2013}, and thus can be used as a standard
candle \citep{Yoo2004}. Figure~\ref{fig:four} shows the locations of
the source and centroid of giant clump in the color-magnitude diagram
of neighboring stars around the source star.  We find that the
de-reddened color and brightness of the source star are
$(V-I,I)_0=(1.04,17.62)$, implying that the source is a K-type
subgiant. From these values, we derive $\theta_*=1.37 \pm 0.10$
$\mu$as by converting $V-I$ into $V-K$ \citep{Bessell1988} and then
applying a color-surface brightness relation \citep{Kervella2004}.

In Table~\ref{table:two}, we list the estimated angular Einstein radii
for both $u_0>0$ and $u_0<0$ solutions. Also presented are the
geocentric and heliocentric lens-source proper motions.  The
geocentric proper motion is determined from the measured angular
Einstein radius and time scale $t_{\rm E}$ by
\begin{equation}
\mu_\oplus={\theta_{\rm E}\over t_{\rm E}}.
\label{eq4}
\end{equation}
With the additional information of $\pivec_{\rm E}$, the heliocentric
proper motion is determined by
\begin{equation}
\muvec_\odot=\muvec_\oplus + {\bf v}_{\oplus,\perp}{\pi_{\rm rel}\over {\rm AU}},
\label{eq5}
\end{equation}
where $\muvec_\oplus=\mu_\oplus(\pi_{{\rm E},N}/\pi_{\rm E},\pi_{{\rm
    E},E}/\pi_{\rm E})$, ${\bf v}_{\oplus,\perp}=(1.3,27.7)\ {\rm
  km}\ {\rm s}^{-1}$ is the velocity of Earth projected onto the sky
at $t_0$, and $\pi_{\rm rel}={\rm AU}(D_{\rm L}^{-1}-D_{\rm S}^{-1})$
is the relative lens-source parallax.

\begin{deluxetable}{lrr}
\tablecaption{Physical parameters\label{table:two}}
\tablewidth{0pt}
\tablehead{
\multicolumn{1}{c}{Quantity}     &
\multicolumn{1}{c}{$u_0>0$}      &
\multicolumn{1}{c}{$u_0<0$} 
}
\startdata
Einstein radius (mas)                      &  1.82 $\pm$ 0.41  &  1.87 $\pm$ 0.43 \\
Geocentric proper motion (mas yr$^{-1}$)   &  7.32 $\pm$ 1.65  &  7.90 $\pm$ 1.82 \\   
Heliocentric proper motion (mas yr$^{-1}$) &  6.16 $\pm$ 1.39  &  6.76 $\pm$ 1.55 \\          
$M_1$ ($M_\odot$)                          &  1.03 $\pm$ 0.24  &  1.03 $\pm$ 0.24 \\ 
$M_2$ ($M_\odot$)                          &  0.91 $\pm$ 0.21  &  0.84 $\pm$ 0.20 \\   
$D_{\rm L}$ (kpc)                          &  3.13 $\pm$ 0.51  &  2.98 $\pm$ 0.50 \\  
$d_\perp$ (AU)                             &  6.11 $\pm$ 0.99  &  6.10 $\pm$ 1.03 \\
$a$ (AU)                                   &  7.6  $\pm$ 4.4   &  10.8 $\pm$ 3.6  \\
$P$ (yr)                                   &  15.4 $\pm$ 13.0  &  23.6 $\pm$ 8.1  \\
Eccentricity                               &  0.36 $\pm$ 0.22  &  0.54 $\pm$ 0.20 \\                   
Inclination (deg)                          & -32.9 $\pm$ 13.3  &  53.6 $\pm$ 5.8  \\                        
Time of perihelion (HJD')                  &  8158 $\pm$ 574   &  8032 $\pm$ 296  
\enddata
\end{deluxetable}

\subsection{Physical Parameters}

With the space-based microlens parallax and the angular Einstein
radius, the mass and distance are estimated by the relations in
Equation (1). We present the determined values in
Table~\ref{table:two} for both $u_0>0$ and $u_0<0$ solutions.  We note
that the two degenerate solutions have similar values of $\pi_{\rm E}$
and $\theta_{\rm E}$ and thus the estimated physical parameters are
similar to each other.  It is found that the binary lens responsible
for OGLE-2015-BLG-0479 is composed of two G-type main-sequence stars
with $M_1\sim 1.0\ M_\odot$ and $M_2\sim 0.9\ M_\odot$ and the
projected separation between the components is $d_\perp\sim 6$ AU.
The estimated distance to the lens is $D_{\rm L}\sim 3$ kpc.

Since we consider a full Keplerian orbital motion, the orbital
parameters are also determined.  The estimated semi-major axis and
orbital period are $a = 7.6\pm 4.4$ AU and $P=15.4 \pm 13.0$ yrs,
respectively, for the $u_0>0$ model and $a = 10.8\pm 3.6$ AU and
$P=23.6 \pm 8.1$ yrs, respectively, for the $u_0<0$ model.  There have
been numerous cases for which the projected orbital parameters
$ds_\perp/dt$ and $d\alpha/dt$ are determined,
e.g. \citet{Albrow2000}. However, it is well recognized that
determining the complete orbital parameters including the
radial-component parameters $s_\parallel$ and $ds_\parallel/dt$ is
very difficult even in very favorable circumstances \citep{Gould2013}
and thus there exist only three cases for which the complete orbital
parameters were measured \citep{Shin2011, Shin2012, Gould2013}.  A
major cause of the difficulty in determining the complete orbital
parameters is the strong correlation between the microlens-parallax
and lens-orbital effects which have similar effects on lensing light
curves. We note that the measurements of the complete orbital
parameters for OGLE-2015-BLG-0479 become possible because the
microlens parallax is precisely measured by the {\it Spitzer}
data. See \citet{Han2016} for detailed discussion about the importance
of space-based microlensing observation in characterizing orbital lens
parameters.

The fact that the lens has a heavier mass than the most common lens
population of low-mass stars and it is located relatively close to the
observer makes us to consider the possibility that the origin of
blended light is likely to be the lens itself.  In order to check this
possibility, we mark the position of the blend in the color-magnitude
diagram presented in Figure~\ref{fig:four}.  We also calculate the
expected position of the lens based on the lens mass (and
corresponding stellar type) and lens distance, as given in
Table~\ref{table:two}.  We first make this calculation under the
assumption that the lens and the clump experience the same extinction
(solid gold point), and then assuming that the lens suffers less
extinction by an amount $0<\Delta A_I<1$ (dashed gold line).  The
slope of the arrow, $R_{VI} = A_I/E(V-I)=1.3$, is determined from the
ratio of total-to-selective extinction along this line of sight toward
the clump.  We note that the blend position is consistent with that
expected for the lens provided the latter lies behind $\Delta
A_I\simeq 0.5$ less extinction than the clump.  That is, the lens
would have to lie behind about 3/4 of the dust.  This is quite
reasonable given the lens distance of $D_L\simeq 3\,$kpc.  The
alternate possibility, i.e., that the blend light comes primarily from
an unrelated star along the line of sight, is virtually ruled out if
the microlens model is correct.  This is because, regardless of how
much dust lies behind the lens, its inferred $I$-band flux already
accounts for the majority of the observed blend light.  Hence, the
room for other, unassociated, stars to contribute to the blend is
highly restricted.  With $I\sim 17.7$, the blend is bright enough for
spectroscopy.  Since the two components of the lens are moving with
internal relative motion of order $\sim 15\,{\rm km\,s^{-1}}$ in both
solutions, the orbit can be measured by making spectroscopic
observations over a number of years.

\section{Summary and Discussion}

We analyzed the combined data obtained from observations both from the
ground and from the {\it Spitzer} telescope for the microlensing event
OGLE-2015-BLG-0479.  The light curves with strong caustic-crossing
features seen from the ground and from space exhibited a time offset
$\sim 13$ days between the caustic spikes, indicating that the
relative lens-source positions seen the two places were displaced by
parallax effects.  From modeling the light curves, we measured the
space-based microlens parallax.  Combined with the angular Einstein
radius measured by analyzing the caustic-crossing parts of the light
curves, we determined the mass and distance of the lens.  It was found
that the lens was a binary composed of two G-type stars with masses
$\sim 1.0\ M_\odot$ and $\sim 0.9\ M_\odot$ located at a distance
$\sim 3$ kpc.  Unlike the binary event OGLE-2014-BLG-1050 observed
also by {\it Spitzer} with similar photometric precision, cadence, and
coverage, we found that interpreting OGLE-2015-BLG-0479 did not suffer
from the degeneracy between $(\pm,\pm)$ and $(\pm,\mp)$ solutions,
confirming that the four-fold parallax degeneracy in single-lens
events collapses into the two-fold degeneracy in general binary-lens
events.  It was found that the location of the blend in the
color-magnitude diagram was consistent with the lens properties,
suggesting that the blend was the lens itself.  The blend is bright
enough for spectroscopy and thus the possibility can be checked from
future follow-up observations.

The binary event OGLE-2015-BLG-0479 analyzed in this work demonstrates
the possibility of characterizing the physical parameters of binary
lenses for a significantly increased number of events. In addition to
the surveys conducted in 2014 and 2015 seasons, the {\it Spitzer}
microlensing survey continues in 2016 season. In addition to {\it
  Spitzer}, the microlensing survey of Campaign 9 of {\it Kepler}'s
extended {\it K2} mission ({\it K2C9}) is being conducted in 2016
season from which microlens parallaxes for $> 127$ microlensing events
are expected to be measured \citep{Henderson2016}. For these
binary-lens events, the chance to measure the angular Einstein radius
is high because of the greatly increased observation cadence of
ground-based surveys achieved by the instrumental upgrade and the
addition of new surveys, e.g. KMTNet survey \citep{Kim2016}. Being
able to measure both $\pi_{\rm E}$ and $\theta_{\rm E}$, therefore, it
will be possible to routinely measure the physical parameters of
binary lenses.

\begin{acknowledgments}
Work by C.~Han was supported by the Creative Research Initiative Program (2009-0081561) of 
National Research Foundation of Korea.  
The OGLE project has received funding from the National Science Centre, Poland, grant 
MAESTRO 2014/14/A/ST9/00121 to AU.  OGLE Team thanks Profs.\ M.~Kubiak and G.~Pietrzy{\'n}ski, 
former members of the OGLE team, for their contribution to the collection of the OGLE 
photometric data over the past years.
Work by AG was supported by JPL grant 1500811.
Work by J.C.Y. was performed under contract with
the California Institute of Technology (Caltech)/Jet Propulsion
Laboratory (JPL) funded by NASA through the Sagan
Fellowship Program executed by the NASA Exoplanet Science
Institute.
Work by CBH and YS was supported by an appointment to the NASA Postdoctoral 
Program at the Jet Propulsion Laboratory, administered by Universities Space 
Research Association through a contract with NASA.
The \textit{Spitzer} Team thanks Christopher S.~Kochanek for graciously
trading us his allocated observing time on the CTIO 1.3m during the
\textit{Spitzer} campaign.
We acknowledge the high-speed internet service (KREONET)
provided by Korea Institute of Science and Technology Information (KISTI).

\end{acknowledgments}

\appendix

\section{On the Issue of False Alarm Probabilities}

\subsection{Naive False Alarm Probabilities}

In principle, one can evaluate the false alarm probabilities (FAPs)
associated
with introducing either $n=2$ or $n=4$ orbital motion parameters
relative to a so-called ``standard'', i.e., non-orbiting binary model,
\begin{equation}
p(n,\Delta\chi^2) \equiv 
[\Gamma(n/2)]^{-1}\int_{\Delta\chi^2/2}^\infty dx x^{n/2-1}e^{-x}
\rightarrow \biggl[\sum_{i=0}^{n/2}{(\Delta\chi^2/2)^i\over i!}\biggr]
\exp(-\Delta\chi^2/2),
\label{eqn:fap}
\end{equation}
where in the last step we have given the explicit expression for even $n$.
To evaluate these FAPs in a conservative fashion, we first renormalize
that $\chi^2$ values in the main text so the $\chi^2/$dof is exactly
unity for the best model, i.e., downward by a factor $666/732.9=0.909$.
Then $\Delta\chi^2=27.3$ and $\Delta\chi^2=39.5$ for $n=2$ and $n=4$,
respectively.  The associated FAPs are then
$p_2(27.3)= 1.1\times 10^{-6}$ and $p_4(39.3)= 6.0\times 10^{-8}$.
These numbers are quite small, and one is tempted to leave it at that.

However, there is actually a deeper issue at stake, which
is that it is fundamentally wrong to evaluate FAPs for this case.
To understand why, we briefly recapitulate a case for which such
evaluation is appropriate.  This will allow us to contrast the key features
of the two cases.

\subsection{Microlens Planet FAPs}

Suppose that a microlensing event is reasonably well fit by a point-lens
model with 3 parameters $(t_0,u_0,t_{\rm E})$ but is better fit by adding
four additional parameters $(s,q,\alpha,\rho)$, with $q\ll1$, indicating
a planet.  Without going into detail (because this is not our main
focus), one can show that the FAP is approximately given by
\begin{equation}
p_{\rm planet}(\Delta\chi^2) \sim 
{1\over f_p}{2 t_{\rm E}\over {t_{{\rm E},p,\min}}}\ln{t_{\rm E}\over 3 t_{{\rm E},p,\min}}
\exp(-\Delta\chi^2/2),
\label{eqn:fap_pl}
\end{equation}
where $f_p\sim 10^{-2}$ is the fraction of all point-lens events with
suitable quality data that show planetary anomalies and
$t_{{\rm E},p,{\rm min}}\sim 1\,$hr is the timescale of the shortest detectable 
planetary anomaly.  The last factor accounts for the $\chi^2$ distribution
associated with 2 additional parameters $(s,\rho)$, while the first
three count the effective number of trials.  The first quantifies
how many events are searched for each real planet.  The second
quantifies the number of independent locations along the light curve
(effectively parameterized by $\alpha$) at which one can
search for planets.  And the third counts the number of independent
durations of this perturbation at fixed location 
(effectively parameterized by $q$).  For typical
Einstein timescales $t_{\rm E}\sim 30\,$days, the first three factors
combine to a value $\sim 10^6$.

Now, such FAPs are never calculated in practice for the simple reason
that no one has ever considered a microlensing planet to be ``detectable''
unless $\Delta\chi^2>160$ \citep{gaudi02}, and in fact all reported
detections have had substantially higher $\Delta\chi^2$.  Even at the
putative threshold of detection, however, the FAP is $\sim 10^{-29}$.
The reasons for this conservative attitude do not concern us here,
but the interested reader can consult \citet{gaudi02} and
\citet{mb11293,mb10311}.

Our focus is rather on a matter of principle.  A planet with mass ratio
$q=0$ yields an absolutely identical model light curve as a point lens.
Hence, if we ``measure'' a mass ratio $q=(1.0\pm 1.0)\times 10^{-4}$,
we do not say that we have ``detected a planet, possibly of zero mass''.
Rather, we formulate this as an upper limit on the mass of any possible planet
that is present (at a given $(s,\alpha)$).  On the other hand,
if the ``measurement'' were $q=(1.0\pm 0.2)\times 10^{-4}$, then
we would think naively that we may have detected a planet and might
ask questions about the FAP.  (As mentioned in the previous paragraph,
no such ``detection'' would ever be considered, but if it were, then
inserting $\Delta\chi^2=25$, one finds that the
FAP would exceed unity!)

\subsection{Orbiting Binary FAPs}

OGLE-2015-BLG-0479 shows two clear caustic crossings, and there
are no known astrophysical phenomena that can generate such
light curve features except having two masses projected on the
sky within of order one Einstein radius of each other.
These two masses must either be bound to each other (and so in a Kepler
orbit) or are unbound, i.e., merely seen in projection,
in which case they are moving relative to each other in rectilinear,
unaccelerated motion.  (The probability of the latter is quite low,
as we discuss immediately below).  In either case, one knows
a priori that they have some instantaneous relative motion.
Hence, one is not ``adding parameters'' to include such motion
($ds/dt$ and $d\alpha/dt$ in the formulations in the main text).
Rather, the opposite is true: if one were to model this -- or any --
binary system (bound or unbound) without including transverse
relative velocity parameters, one would be suppressing the
impact of known physics on the light curve and so, possibly,
introducing systematic errors on the remaining parameters being
measured.

In particular, if this measurement showed a best fit of zero
relative motion (or consistent with zero motion at low
$\Delta\chi^2$), we would not say (as with $q\simeq 0$ in the planet
case) that we had failed to measure transverse motion.  Still
less would we say that there was ``no justification'' for introducing
transverse-velocity parameters.  Rather, we would say that we
had {\it measured} the transverse velocity to be close to zero.
And this measurement would be quite important because it would
provide additional evidence that the system is bound.

That is, the probability of finding two unrelated stars projected
within about one Einstein radius is already low, but the probability
that they are moving slowly with respect to each other is yet
another factor $\sim 100$ lower.  The first probability is roughly the
optical depth to microlensing, i.e., $\tau\sim 10^{-6}$.  This
might seem too small to consider, but since there have been more
than 20,000 microlensing events discovered to date, the probability
of such a chance projection in one of these is a few percent.
Hence, the additional suppression factor of $\sim 100$ from
measuring a small transverse motion can be important.

Once the binary is demonstrated to be bound with very high probability,
it is certainly justified to ``introduce'' the remaining two parameters
needed to describe a full Kepler orbit.  We have put ``introduce''
in quotes because nothing is being introduced: rather we are simply
not eliminating parameters that are known to be required to describe
the physical system.

\subsection{When Is One Justified in Eliminating Some or All Kepler Parameters?}

 From a purist standpoint, the answer is never.  The practical reason
that these ``extra'' parameters are frequently excluded is that in many cases
nothing would be measured by doing so.  For example, in many cases one
finds that the
transverse motion is consistent with zero but is equally consistent
with values several times higher than permitted for bound orbits.
Since, as just mentioned, the prior probability for bound orbits is
quite high, such a ``measurement'' yields no information.   One
is then tempted to simply set this motion to zero, i.e., fit the
data without these two parameters.  And for many years this is
exactly what was done.  However, such an approach is unphysical:
binary stars do not ``stand still''.  Furthermore, as first shown
by \citet{mb09387} and then further elaborated by \citet{Skowron2011},
if the transverse motion parameters are arbitrarily set to zero, then this
can introduce systematic errors into the parallax parameters, with
which they are correlated.  Rather the correct approach is to maintain
these parameters.  Then if they take on improbable or unphysical
values, the proper way to handle this is to introduce Bayesian
priors on these parameters.  See, for example, \citet{ob120406}.
On the other hand, if the light curve does not contain enough information
to fruitfully constrain either the transverse motion parameters or
the parallax parameters, then setting these to zero is in most cases
an appropriate way to simplify the fitting, since the remaining
parameters are usually not strongly correlated with them.

\end{document}